\documentclass{PoS}

\newcommand{\sign}{\mbox{sign}}

\title{Lattice simulation of SU(2) gauge theory with chirally symmetric fermions}

\ShortTitle{Lattice simulation of SU(2) gauge theory with chirally symmetric fermions}

\author{\speaker{Hideo Matsufuru} \\%
 High Energy Accelerator Research Organization (KEK), Tsukuba 305-0801, Japan\\
 E-mail: \email{hideo.matsufuru@kek.jp}
}

\author{Yoshio Kikukawa\\
Institute of Physics, University of Tokyo, Tokyo 153-8092, Japan\\
 E-mail: \email{kikukawa@help.c.u-tokyo.ac.jp}
}

\author{Kei-ichi Nagai\\
 Kobayashi-Maskawa Institute for the Origin of Particles and the Universe (KMI),
  Nagoya University, Nagoya, 464-8602, Japan\\
 E-mail: \email{keiichi.nagai@kmi.nagoya-u.ac.jp}
}

\author{Norikazu Yamada\\
 High Energy Accelerator Research Organization (KEK), Tsukuba 305-0801, Japan,\\
 and Graduate University for Advanced Studies (Sokendai),
 Tsukuba 305-0801, Japan
\\
 E-mail: \email{norikazu.yamada@kek.jp}
}

\abstract{
We numerically study the SU(2) gauge theory with two dynamical
flavors of the domain-wall fermions in fundamental representation.
The meson spectra and the residual mass are measured on three
lattice volumes and at two values of gauge coupling so as to
investigate the finite volume effect.
On generated configurations, eigenvalues of the overlap fermion
operator are determined and compared to the random matrix theory.
To quantify the effect of violation of the exact chiral symmetry,
we measure the correlation between the eigenvectors of
the domain-wall and the overlap operators.
}

\FullConference{
31st International Symposium on Lattice Field Theory LATTICE 2013\\
July 29 - August 3, 2013\\
Mainz, Germany}

\begin{document}

\section{Introduction}

The SU(2) gauge theory has for long time been studied as a prototype
theory of QCD that shares important features with QCD such as chiral
symmetry breaking and confinement, while with less numerical cost.
Recently it has been drawn much attention in search for a theory
beyond the standard model as a candidate of technicolor theory
\cite{Neil:2012cb,Lewis:2011zb,Karavirta:2011zg,Hayakawa:2013yfa}.
In this work, we also focus on the chiral dynamics of SU(2) gauge
theory.
The pattern of spontaneous chiral symmetry breaking depends on the
gauge group and the fermion representation.
QCD, and in general SU($N$) with $N\geq 3$ gauge theory with
fundamental fermions are considered to chirally break down as 
$SU(N_f)\times SU(N_f)\rightarrow SU(N_f)$, where $N_f$ is the number
of flavors.
Since SU(2) group is pseudoreal, it has a pattern of chiral symmetry
breaking different from QCD: SU(2$N_f$)$\rightarrow$Sp(2$N_f$).
The adjoint representation of fermions results in another pattern
SU(2$N_f$)$\rightarrow$SO(2$N_f$) for any $N$.
One of the motivations of this work is to understand the effect of
the symmetry breaking pattern on the chiral dynamics,
in particular its dependence on the number of flavors and
temperature.

To explore the chiral symmetry on the lattice, it is important to 
employ the fermion formulation that retains the chiral symmetry
as exact as possible.
In this sense, the overlap fermion operator,
\begin{equation}
 D_{ov} = M_0 \left[ 1 + \gamma_5 \sign H_W(-M_0) \right] ,
\label{eq:overlap_operator}
\end{equation}
is the best solution, where $H_W$ is the Hermitian Wilson-Dirac
operator with $M_0$ the domain-wall height.
$D_{ov}$ satisfies the Ginsparg-Wilson relation, which expresses
an exact chiral symmetry on the lattice.
In particular, simulations in the so-called $\epsilon$-regime
are an attractive device to extract the chiral condensate precisely.
However, its numerical cost is high and to reduce it down to
an acceptable level, elaborated setup, such as reduction of
near-zero modes of $H_W$ by employing the topology fixing term,
is demanded.
Such situation may cause the simulation and analysis involved.

An alternative approach is the domain-wall (DW) fermion.
The DW fermion operator is defined as a massive fermion
in the five-dimensional space, and light modes appear on the
boundaries in the 5th dimension.
The DW action is written as
\begin{eqnarray}
 S_{DW} &=& \sum_{x,s} \bar{\psi}(x,s) D_W(x,y;-M_0) \psi(y,s)
\nonumber \\
 & &
 - \frac{1}{2}\sum_{x,s} \bar{\psi}(x,s) \left[
    (1-\gamma_5) \psi(x,s+1) + (1+\gamma_5) \psi(x,s-1)
   -2\psi(x,s) \right]
\nonumber \\
 & & + m \left[ \bar{\psi}(x,1) P_R \psi(x,L_s) +
                \bar{\psi}(x,L_s) P_L \psi(x,1) \right] ,
\label{eq:domain-wall_action}
\end{eqnarray}
where $m$ is fermion mass, $P_{R,L} = (1 \pm \gamma_5)/2$,
$L_s$ the size of 5th dimension, and $D_W$ the Wilson
operator with negative mass,
\begin{equation}
  D_W(x,y;-M_0) = - M_0 \delta_{x,y}
   - \frac{1}{2}\sum_{\mu=1}^4 \left\{
      (1-\gamma_\mu) U_\mu(x)\delta_{x+\hat{\mu},y}
    + (1+\gamma_\mu) U_\mu^\dag (x-\hat{\mu}) \delta_{x-\hat{\mu},y}
    - 4 \delta_{x,y}  \right\} .
\end{equation}
Taking the limit of $L_s\rightarrow \infty$ with fixed
$a_s L_s$, the action (\ref{eq:domain-wall_action}) accompanied by
Pauli-Villars ghost term arrives at the overlap fermion operator
(\ref{eq:overlap_operator}).
How much the chiral symmetry is violated in DW fermion action
is probed by a quantity
\begin{equation}
 R(t) = \frac{\sum_{\vec{x}} \langle J_{5q} (\vec{x},t) P(0) \rangle}
             {\sum_{\vec{x}} \langle P(\vec{x},t) P(0) \rangle} ,
\label{eq:residual_mass}
\end{equation}
where
\begin{equation}
 J_{5q}(x) = - \bar{\psi}(x,L_s/2) P_L \psi(x,L_s/2+1)
          + \bar{\psi}(x,L_s/2+1) P_R \psi(x,L_s/2) .
\end{equation}
An average of $R(t)$ over large $t$ separation is a standard definition
of the residual mass.

Our original strategy is to explore the dynamics of the chiral symmetry
breaking with the overlap fermion in the $\epsilon$-regime in which
the Compton wave length of the PS meson is larger than the system size
\cite{Matsufuru:2010zz}.
The eigenvalue distribution of the overlap fermion can be compared to
the chiral random matrix theory so as to extract the chiral condensate
with high precision.
However, due to aforementioned complication, we decided to perform
first the dynamical simulations with DW fermions as a reference and for
systematic survey.

In this paper, we focus on the SU(2) gauge theory with two flavors of
fundamental domain-wall fermions, as a basis to investigate the $N_f$
dependence and the properties at finite temperature.
The setup of numerical simulations is described in the next section.
Sections~\ref{sec:domainwall} and \ref{sec:overlap} describe
the results of measurement with valence domain-wall and overlap
operators, respectively.
How the eigenmodes of the former approach to the latter is
discussed in Sec.~\ref{sec:correlation}.
The last section is devoted to summary and outlook.

\section{Simulation setup}
\label{sec:setup}

\begin{figure}[tb]
\begin{center}
\includegraphics[clip=true,width=0.5\textwidth]{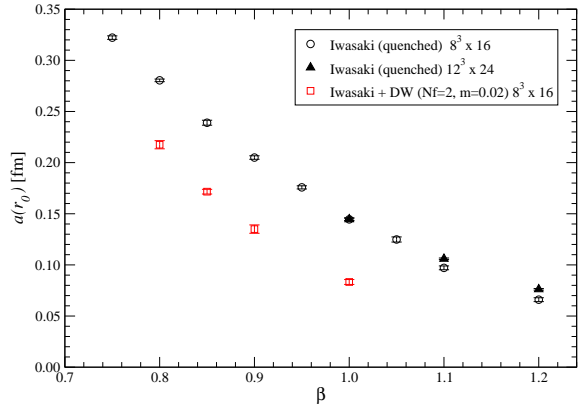}
\vspace{-0.6cm}
\end{center}
\caption{
$\beta$-dependence of the lattice spacing defined through the
Sommer's hadronic radius for quenched and $N_f=2$ simulations
with the Iwasaki gauge and the domain-wall fermion actions.
}
\label{fig:scale} 
\end{figure}

The gauge configurations are generated with the Iwasaki gauge action
with $N_f=2$ dynamical fermions with the standard DW fermions
accompanied by the Pauli-Villars ghosts.
We employ the Hybrid Monte Carlo algorithm with the Omelyan
integrator and multi-time-step.
The code is implemented by modifying the Bridge++ code set
\cite{Bridge++} so as to enable SU(2) simulations.
The lattice sizes are $8^3\times 16$, $12^3\times 24$, and $16^3\times 32$.
As the parameters of the DW fermions, we use $L_s=16$, and $M_0=1.6$
throughout present work.

We first observe the $\beta$-dependence of the lattice spacing.
Although the lattice spacing in physical units does not make sense for the
present theory, it is often convenient to draw a rough idea of the scale.
Figure~\ref{fig:scale} displays the $\beta$-dependence of the lattice
spacing defined through the Sommer's hadronic radius $r_0=0.05$fm
that was determined from the static fermion potential in the quenched
and $N_f=2$ cases.
Based on this result, we decided to adopt $\beta = 0.85$ and 0.90
as our first target of numerical simulations.
The mass of the dynamical domain-wall fermion is varied from $m=0.02$
to 0.20.

\section{Meson spectrum and residual mass}
\label{sec:domainwall}

\begin{figure}[tb]
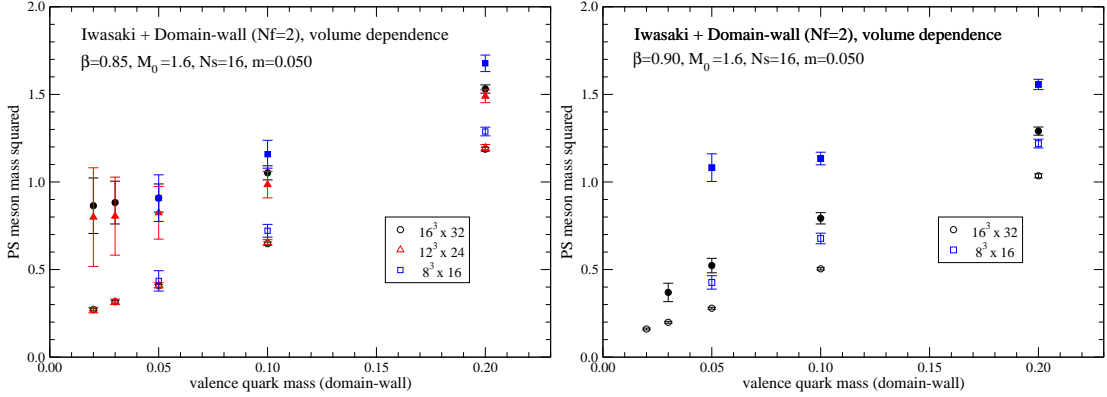

\begin{center}
\includegraphics[clip=true,width=0.48\textwidth]{Figs/mass2_RgDW_Vdep_b0.85_Ns16_M01.60_Nf02_mq0.050.eps}
\includegraphics[clip=true,width=0.48\textwidth]{Figs/mass2_RgDW_Vdep_b0.90_Ns16_M01.60_Nf02_mq0.050.eps}
\vspace{-0.6cm}
\end{center}
\caption{
The finite size effect on the masses of PS and V mesons
at $\beta=0.85$ (left panel) and 0.90 (right) and the sea fermion
mass $m=0.050$.
}
\label{fig:finitesize_effect}
\end{figure}

We first measure the meson spectrum and the residual mass
under variation of valence fermion mass.
Using a local meson operator at both the source and sink,
statistics of 20-40 configurations is sufficient to determine
the PS meson masses, while not for the other channels.
Figure~\ref{fig:finitesize_effect} shows the volume dependence
of the PS and V meson masses at $\beta=0.85$ and 0.90 with
$m=0.05$.
The PS meson mass is proportional to the valence fermion mass,
as is familiar in QCD.
The finite intercept indicates the explicit breaking of the chiral
symmetry in the DW fermion formulation.
The offset values are consistent with the direct calculation of
the residual mass with Eq.~(\ref{eq:residual_mass}).
While the values of the residual mass is not negligibly small,
it decreases as $\beta$ increases as is expected.

\begin{figure}[tb]
\begin{center}
\includegraphics[clip=true,width=0.5\textwidth]{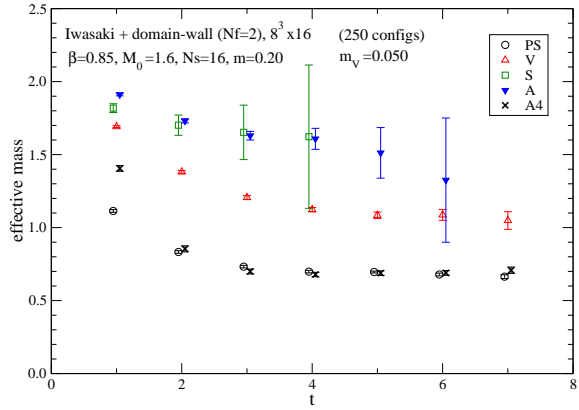}
\vspace{-0.6cm}
\end{center}
\caption{
Meson masses in PS, V, S and A channels for the valence fermion mass
$m_V=0.050$ on 250 configurations generated at $\beta=0.85$ and
the sea fermion mass $m=0.20$ on the $8^3\times 16$ lattice.
}
\label{fig:other_channels}
\end{figure}

The left panel of Fig.~\ref{fig:finitesize_effect} compares the
meson masses in three volumes at $\beta=0.85$.
While the values for $L=12$ and $16$ are consistent, the result for $L=8$
is slightly larger.
Extrapolating to $m_V=0$, $Lm_{PS}\simeq 5.4$ for $L=12$ and 3.6 for $L=8$,
and thus the above results are consistent with the wisdom in QCD that
$Lm_{PS}>4$ is required to avoid large finite volume effect.
At $\beta=0.90$ displayed in the right panel of
Fig.~\ref{fig:finitesize_effect}, only $L=8$ and $16$ cases have been
obtained which are significantly deviate.
Since $Lm_{PS}\simeq 5.0$ for $L=16$ and 3.6 for $L=8$,
to settle the finite volume effect we need a result for $L=12$ that
computation is now in progress.

Once the finite size effect and size of chiral symmetry breaking are
quantified, we can systematically investigate the meson spectrum and
other observables.
For the channels other than PS and V, however, the correlators suffer
from larger statistical fluctuation.
In Figure~\ref{fig:other_channels} we show an example of meson masses
on $8^3\times 16$ lattice at $\beta=0.85$ with higher statistics.
By applying the same setup as PS and V mesons, $O(250)$ configurations
are not enough to extract the masses of S and A channels in sufficient
precision.
Some kind of techniques to increase the signal is necessary such as
the all-to-all propagators.

\section{Valence overlap fermion operator}
\label{sec:overlap}

\begin{figure}[tb]
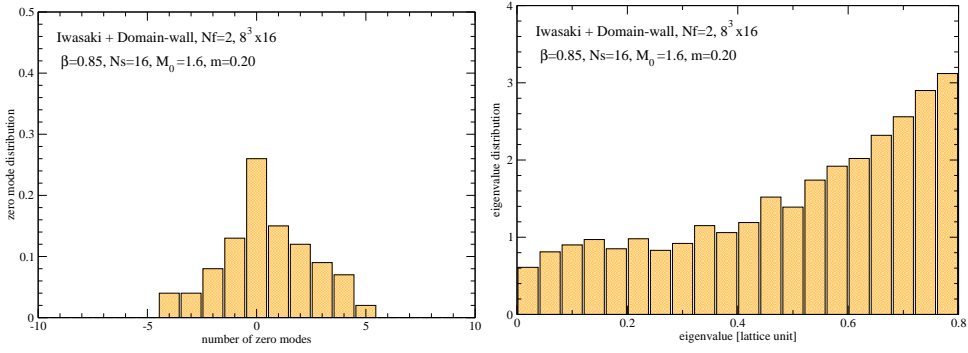

\begin{center}
\includegraphics[clip=true,width=0.42\textwidth]{Figs/zero_dist_RgDW_08x16_b0.85_Ns16_M01.60_Nf02_mq0.200.eps}
\includegraphics[clip=true,width=0.42\textwidth]{Figs/eigen_dist_RgDW_08x16_b0.85_Ns16_M01.60_Nf02_mq0.200.eps}
\vspace{-0.6cm}
\end{center}
\caption{
The distribution of zero modes (left panel) and the density of
low-lying eigenvalues (right) of the valence overlap fermion operator
at $\beta=0.85$ and $m=0.20$ on $8^3\times 16$ lattice.
}
\label{fig:zero_mode}
\end{figure}

It is interesting to measure the eigenmodes of the valence
overlap fermion operator on the configurations generated with
DW fermions.
As numerical implementation of the overlap operator
(\ref{eq:overlap_operator}),
we adopt the Zolotarev's partial fractional approximation to
the sign function with the low-mode projection of $H_W$.
The low-lying eigenmodes of massless $H_{ov}= \gamma_5 D_{ov}$ are
determined by applying the implicitly restarted Lanczos algorithm.
Figure~\ref{fig:zero_mode} shows an example of measurement
on $8^3\times 16$ lattice at $\beta =0.85$ and $m=0.20$.
The left panel shows the distribution of the number of
zero modes whose deviation give the topological susceptibility.
The right panel shows the density of the low-lying eigenmodes.
Clearly the result shows the expected behavior for the broken
chiral symmetry through the Banks-Casher relation.
Quantitative analysis is now underway.

Once the distribution of the eigenvalues are determined, it
can be compared to the prediction of the random matrix theory.
In this context the most interesting quantity is the distribution
of low-lying eigenvalues, that is to be compared with the
chiral random matrix theory \cite{Damgaard:2000ah}.
However, our present setup and statistics are not suitable to
such an analysis.
We instead compare the level statistics of the bulk region of
the eigenvalue spectrum to the random matrix theory.
It is not related to the chiral dynamics but just reflects the
symmetry of the fermion operator.

\begin{figure}[tb]
\begin{center}
\includegraphics[clip=true,width=0.5\textwidth]{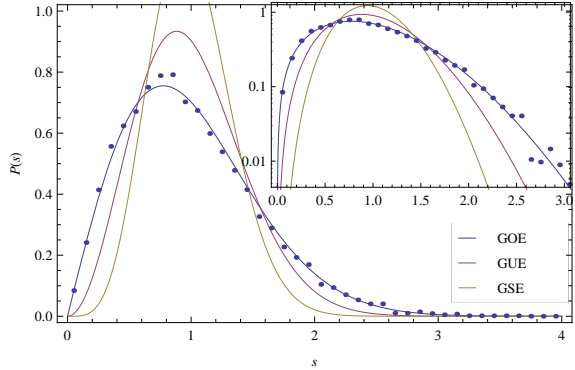}
\vspace{-0.8cm}
\end{center}
\caption{
Unfolded level spacing of the overlap fermion operator at $\beta=0.85$
and $m=0.20$ on $8^3\times 16$ lattice.
The predictions of the random matrix theory for GOE, GUE, and GSE are
also displayed.
}
\label{fig:random_matrix_theory}
\end{figure}

Figure~\ref{fig:random_matrix_theory} shows the density of the
unfolded level spacing at $\beta=0.85$ and $m=0.20$%
\footnote{
This analysis was performed by S.~M.~Nishigaki.
}.
The unfolding is applied to the eigenmodes in the range
[0.2,0.8] which are measured on 500 configurations irrespective
to the topological charge.
In the figure, predictions of the random matrix theory with
three types of ensembles,
orthogonal (GOE), unitary (GUE), and symplectic (GSE) are 
also displayed.
The numerical result is well consistent with the GOE curve
that corresponds to the symmetry of fundamental representation
of SU(2) group.

\section{Correlation between DW and overlap eigenmodes}
\label{sec:correlation}

Since the dynamics of the chiral symmetry breaking is encoded
in the low-lying eigenmodes, these modes can be used to quantify
the effect of chiral symmetry violation brought by adopting
the domain-wall fermion action.
The four-dimensional domain-wall operator
\cite{Kikukawa:1999sy,Borici:1999zw},
\begin{equation}
 D^{(4)}(m) = \{ P^{-1} [D_{DW}(1)]^{-1} D_{DW}(m) P \}_{11}
\label{eq:4D-domainwall}
\end{equation}
where $P_{ij}=\delta_{j,\mbox{\scriptsize mod}(i+1,L_s)}$,
is the counterpart of the overlap operator that approaches
Eq.~(\ref{eq:overlap_operator}) in the limit $L_s \rightarrow \infty$.
Thus we investigate the correlation between the eigenvectors
of Eq.~(\ref{eq:overlap_operator}) and of
Eq.~(\ref{eq:4D-domainwall})  determined on the same configurations.

\begin{figure}[tb]
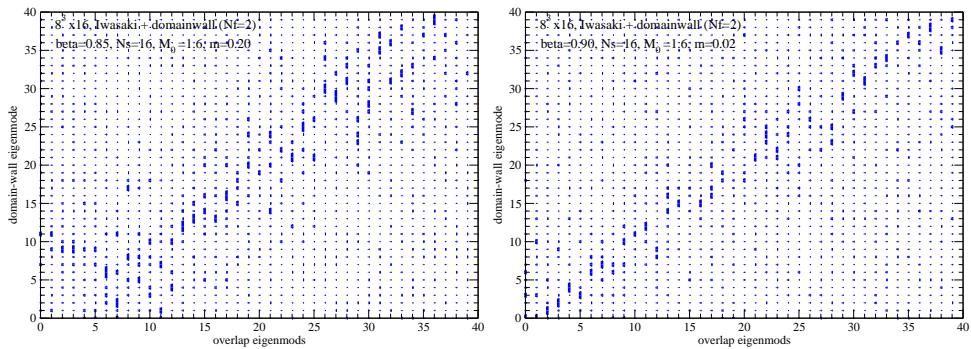

\begin{center}
\includegraphics[clip=true,width=0.42\textwidth]{Figs/interfere_RgDW_08x16_b0.85_Ns16_M01.60_Nf02_mq0.200_conf0320.eps}
\includegraphics[clip=true,width=0.42\textwidth]{Figs/interfere_RgDW_08x16_b0.90_Ns16_M01.60_Nf02_mq0.020_conf0610.eps}
\vspace{-0.6cm}
\end{center}
\caption{
Correlation between the eigenmodes of the domain-wall and overlap operators.
The left and right panels are at $\beta =0.85$ and 0.90, respectively.
}
\label{fig:correlation_eigenmodes}
\end{figure}

Figure~\ref{fig:correlation_eigenmodes} displays the correlation of
the 40 lowest eigenmodes of the 4D domain-wall and overlap fermion
operators, $C_{i,j} = v^{(DW)\dag}_i \cdot v^{(ov)}_j$,
on single configuration.
The left and right panels show the results at $\beta =0.85$
($m_{res}\simeq 0.03$) and $\beta =0.90$ ($m_{res}\simeq 0.015$),
respectively.
As expected, near-diagonal components of the correlation increase
as the residual mass decreases.
The same tendency is generally observed on other configurations,
while a quantitative analysis is underway.

\section{Conclusion and outlook}
\label{sec:conclusion}

We are investigating the SU(2) gauge theory with dynamical fermions
in the fundamental representation.
Numerical simulations with two-flavors of domain-wall fermions
were performed as a basis for extensive studies with variation
of number of flavors and at finite temperature.
The results obtained are feasible for further investigation,
while techniques to increase the signal is required for quantitative
computation of variety of meson sectors.
We also observed the eigenmodes of the valence overlap operator
which is useful to explore the effect of the chiral symmetry 
violation of the domain-wall fermions.
We are preparing for systematic studies of subjects described in
this paper, as well as for applying the same procedure to
the adjoint representation of fermions.

\section*{Acknowledgment}

We thank S.~M.~Nishigaki for useful discussion and for analysis to
generate Fig.~\ref{fig:random_matrix_theory}.
Numerical simulations were performed on Hitachi SR16000 and IBM Blue Gene/Q
at KEK under a support of its Large-scale Simulation Program
(No.12/13-15) and $\varphi$ computer system at KMI, Nagoya University.
We also thank the Japan Lattice Data Grid which is a grid file system
constructed on a virtual private network SINET4 provided by National
Institute of Informatics for efficient data transfer.
This work is supported in part by the Grand-in-Aid for
Scientific Research of the Japan
(Nos.20105005, 22224003, 22740183, 25400284).

\end{document}